\documentclass[conference]{IEEEtran}
\IEEEoverridecommandlockouts
\usepackage{cite}
\usepackage{amsmath,amssymb,amsfonts}
\usepackage{algorithmic}
\usepackage{graphicx}
\usepackage{textcomp}
\usepackage{xcolor}
\def\BibTeX{{\rm B\kern-.05em{\sc i\kern-.025em b}\kern-.08em
    T\kern-.1667em\lower.7ex\hbox{E}\kern-.125emX}}

\usepackage[hyphens]{url}
\usepackage{subcaption}

\usepackage{tabularx,booktabs}
\newcolumntype{C}{>{\centering\arraybackslash}X} 
\usepackage{lipsum} % filler text
\usepackage{caption}
\usepackage[numbers]{natbib}
\usepackage{threeparttable}
\usepackage{amsmath}

\usepackage{hyperref}

\hypersetup{
    colorlinks=true,
    linkcolor=black,   % Color for internal links
    citecolor=black,   % Color for citation links
    urlcolor=black     % Color for URLs
}

\begin{document}

%\title{Empowering Low-Resource Validator Nodes in Blockchain Networks: A Scalable State Sharing Protocol}

\title{A Scalable State Sharing Protocol for Low-Resource Validator Nodes in Blockchain Networks}

\author{\IEEEauthorblockN{Ruben Hias}
\IEEEauthorblockA{\textit{KU Leuven} \\
Leuven, Belgium \\
ruben.hias@student.kuleuven.be}
\and
\IEEEauthorblockN{Weihong Wang}
\IEEEauthorblockA{\textit{DistriNet, KU Leuven} \\
Leuven, Belgium \\
weihong.wang@kuleuven.be}
\and
\IEEEauthorblockN{Jan Vanhoof}
\IEEEauthorblockA{\textit{DistriNet, KU Leuven} \\
Leuven, Belgium \\
jan.vanhoof1@kuleuven.be}
\and
\IEEEauthorblockN{Tom Van Cutsem}
\IEEEauthorblockA{\textit{DistriNet, KU Leuven} \\
Leuven, Belgium \\
tom.vancutsem@kuleuven.be}
}

\maketitle

% TODO: Remove it. Just for showing pages
\thispagestyle{plain}
\pagestyle{plain}

\begin{abstract}
The perpetual growth of data stored on popular blockchains such as Ethereum leads to significant scalability challenges and substantial storage costs for operators of full nodes. Increasing costs may lead to fewer independently operated nodes in the network, which poses risks to decentralization (and hence network security), but also pushes decentralized app developers towards centrally hosted API services.

This paper introduces a new protocol that allows validator nodes to participate in a blockchain network without the need to store the full state of the network on each node. The key idea is to use the blockchain network as both a replicated state machine and as a distributed storage system. By distributing states across nodes and enabling efficient data retrieval through a Kademlia-inspired routing protocol, we reduce storage costs for validators. Cryptographic proofs (such as Merkle proofs) are used to allow nodes to verify data stored by other nodes without having to trust those nodes directly. While the protocol trades off data storage for increased network bandwidth, we show how gossiping and caching can minimize the increased bandwidth needs.

To validate our state sharing protocol, we conduct an extensive quantitative analysis of Ethereum's data storage and data access patterns. Our findings indicate that while our protocol significantly lowers storage needs, it comes with an increased bandwidth usage ranging from 1.5 MB to 5 MB per block, translating to an additional monthly bandwidth of 319 GB to 1,065 GB. Despite this, the size remains small enough such that it can be passed to all nodes and validated within Ethereum's 12-second block validation window. Further analysis shows that Merkle proofs are the most significant contributor to the additional bandwidth. To address this concern, we also analyze the impact of switching to the more space-efficient Verkle Proofs. Our findings show that Verkle Proofs would promise significant efficiency gains if integrated into the protocol. \end{abstract}


\section{Introduction} \label{sec:intro}

A blockchain is a decentralized digital ledger where every transaction is permanently recorded. This enables decentralized network functionality in adversarial contexts but also implies that the ledger's size continuously increases, which is unsustainable in the long run.

Consider a widely-used permissionless network such as Ethereum. At the time of writing, an Ethereum full node requires at least 2TB of SSD storage~\cite{HardwareRequirements}, a significant investment for individual contributors. Without these resources, they may turn to managed node providers, with a cost of about USD 240 per month~\cite{AllnodesPricing}. As these costs continue to rise, smaller parties are discouraged from operating full nodes which introduces various risks.

The first risk is the \textbf{security and resilience} of the network. Fewer nodes imply fewer points of failure, reducing network resilience. For instance, the smaller the number of network nodes, the easier it is to stage eclipse attacks, where an attacker isolates a target node to feed it false information~\cite{heilmanEclipseAttacksBitcoin2015}.

The second risk is \textbf{network congestion}. Fewer nodes to handle transaction loads can lead to slower confirmations and higher gas fees.

The third risk is \textbf{centralization}. With the Proof-of-Stake model, multiple logical validators can use a single physical node, enabling cheap staking services. For example, Lido controls nearly a third of all staked Ethereum, with around 9,366,644 Ether valued at 27 billion dollars as of May 9th 2024~\cite{StakeLidoLido, etherscan.ioStETHStETHToken, StakedETHChart}.

% Another domain where centralization occurs is in decentralized applications,

The high cost of running a full node also results in \textbf{the reliance on centralized gateways} for decentralized applications. These applications
often run in browsers and need to get their blockchain data from a node. The cost
of running a node for these applications is often prohibitively high, so
decentralized application developers often opt for a centralized Node-as-a-Service (NaaS), such as Infura~\cite{Web3DevelopmentPlatform}, where they can send API requests to and pay for their usage. 

% The problem with this approach is that the application has no way to validate the responses of these providers.

The potential reluctance of users with limited resources to join the
network poses a significant threat to the long-term maintenance and
decentralization of blockchain systems. To address this problem, it is crucial
to lower the cost of operating a node. The main challenge is how to deal with rising storage costs as the size of the state and block history will continue to grow over time.

This paper proposes a new way of handling the ever-increasing state, without changing the functioning of the blockchain. This is achieved by using the blockchain network not only as a replicated state machine but also as a distributed storage network. In most blockchains, each node is required to have the state of all accounts available to be able to process incoming transactions. Our method has the state distributed over all the nodes in the network, with a certain replication factor. This enables individual nodes to retrieve the necessary data on-demand, validate its correctness, and then use the data to validate and process new transactions.
% \textcolor{red}{TODO: First introduce the analysis, then talk about the protocol, in order to align with the rest of our paper.}

Our system achieves this through a two-fold approach: Firstly, it provides a protocol inspired by Kademlia~\cite{maymounkovKademliaPeertoPeerInformation2002} that enables nodes to independently determine whether they need to store a portion of the state and efficiently locate state information that is not stored locally. Secondly, it implements an optimized gossiping and caching method to minimize the additional bandwidth consumed during this process. By combining these two components, the system ensures efficient state management and reduces the overall bandwidth usage. Even with a moderate cache size of $\sim$100 MB, the needed bandwidth can be cut in half with respect to a naive approach.

To benchmark our system and to optimize it for realistic data access patterns of blockchain transactions, we conduct an extensive quantitative analysis of the Ethereum network. We use a combination of data sources (Google BigQuery Ethereum dataset~\cite{EthereumBigQueryPublic}, Paradigm Data Portal~\cite{ParadigmDataPortal} and data requested through an Ethereum full node) with each result calculated over a time span of at least two weeks.

% To benchmark our system, we conducted an analysis of which data is accessed for each transaction on the Ethereum network. 

The study of access patterns in Ethereum yields interesting results that we believe were previously overlooked and that may prove useful beyond our work.
Key findings include a significant $\sim$6x discrepancy between the average code size and the average size of accessed code. Additionally, we observe a 90\% reduction in storage by de-duplicating code with the same storage hashes. Our research also provides an in-depth examination of Merkle Proofs and Verkle block witness sizes, highlighting the potential improvements in block size by switching to Verkle Proofs.

The structure of this paper is as follows: Section \ref{sec:bg} provides a comprehensive background on Kademlia. In Section \ref{ethereum-access-patterns}, we present a detailed quantitative analysis of Ethereum. The proposed state network protocol is outlined in Section \ref{sec:statenetwork}, followed by an evaluation of its performance in Section \ref{sec:eval}. Section \ref{sec:relatedwork} reviews related work in the field, and Section \ref{sec:futurework} discusses the limitations of this study and future research.

\section{Background} \label{sec:bg}
Our state sharing protocol leverages an adaptation of Kademlia for routing requests to find and retrieve data. Understanding Kademlia's principles and structure helps illuminate the mechanisms that enable efficient and reliable data retrieval in our proposed protocol.

\subsection{Kademlia Distributed Hash Table}\label{kademlia}

Kademlia~\cite{maymounkovKademliaPeertoPeerInformation2002} is a distributed
hash table used for routing, featuring several properties that make it appealing
for decentralized networks.

% In Kademlia, each piece of content and each node has its own ID, traditionally 160 bits in
% length. Each node maintains 160 k-buckets, which are used to route requests and
% locate nodes that store specific pieces of content. 

Each piece of content and node has a 160-bit ID. Nodes maintain 160 k-buckets for routing and locating specific content. This matches with the tree structure often adopted in the state storage of blockchains. When a node receives a request or reply, it adds the node triplet (IP address, port, node ID) to a
k-bucket based on the external node's ID.

\begin{figure}[h]
    \centering
    \includegraphics[width=0.4\textwidth]{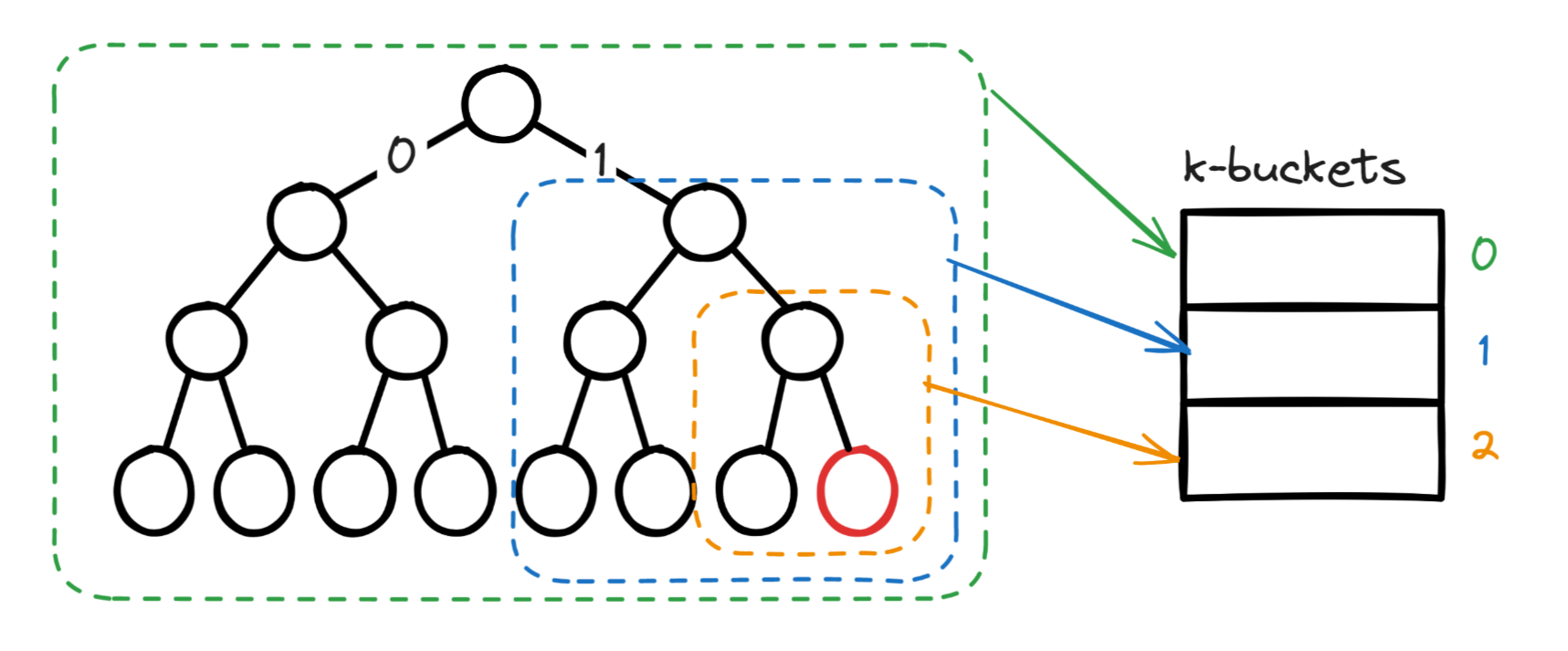}
    \caption{Bucket distribution in Kademlia}
    \label{fig:bucket-distribution-kademlia}
\end{figure}

The triplet is added to the bucket corresponding to the number of \emph{common starting bits} between the node's own ID and the external node ID. For example, if the first two bits are the same but the third differs, the triplet will be placed in the third k-bucket ($k = 2$), as shown in Figure \ref{fig:bucket-distribution-kademlia}. This system means that each subsequent bucket has
exponentially fewer potential nodes that fit into it. As a result, each
node has more detailed knowledge of nodes ``close'' to them due to the increased
number of buckets available for nearby nodes.

The eviction mechanism of k-buckets maximizes node availability. When a request or reply comes in from a node already present in the k-buckets, its triplet is moved to the bucket's tail. If the node is new, three possible scenarios can occur. First, if the bucket is not yet full, the triplet is added to the tail. Second, if the bucket is full, the node will ping the oldest node at the head of the bucket to check its liveliness. If the check fails, it is evicted, and the new triplet is added to the tail. Third, if the pinged node responds, it is moved to the tail, and the new node is discarded. This mechanism optimizes node uptime and future liveliness.

Kademlia uses an iterative lookup procedure starting with the requesting node looking up the closest nodes to the content ID, using the XOR of the addresses as a distance metric. It makes parallel requests to these nodes for the data. If a node has the value, it returns it, and the process stops. Otherwise, the nodes return the closest nodes in their k-buckets. The k-bucket structure gives each node more knowledge of the network ``closer'' to itself, making the iterative process effective.

\section{Quantitative Analysis of Ethereum}\label{ethereum-access-patterns}

Designing a high-performance distributed storage system for a smart-contract-enabled blockchain network requires a deep understanding of the network's data access patterns.

This section analyzes Ethereum's smart contracts and storage, focusing on account headers, code, storage, and proofs. By understanding these patterns, we can optimize the protocol for efficient retrieval and minimal latency.

\subsection{Data Sources}

Three data sources were used for the statistics in this section. The first is the public Google BigQuery ``Blockchain Analytics Ethereum Mainnet" dataset~\cite{EthereumBigQueryPublic}. The second is the open-source \emph{slots} dataset from the Paradigm Data Portal~\cite{ParadigmDataPortal}. The third is Chainstack~\cite{FastReliableBlockchain}, which provides Ethereum nodes for sending JSON-RPC requests, filling gaps in existing datasets. Each statistic covers a period of at least 2 weeks.

\subsection{Account Locality}

Transactions in Ethereum exhibit high locality, meaning they tend to access the
same set of accounts repeatedly. This behavior improves efficiency since
frequently accessed accounts can be cached, reducing the need for repeated data
transfer and lowering overall network load. For example, the top 100 accounts contribute 45\% of all account accesses while only representing 0.0015\% of all observed accounts. Two-thirds of all accounts accessed during the observed time period only got accessed once or twice.

\subsection{Account Size}
To determine the additional bandwidth required for state transfer, we must calculate the size of an account, which includes both the account header and additional data required to execute contracts. The main storage structure in Ethereum, the State Trie, stores all account headers containing basic information such as address, nonce, and balance. External account headers total 60 bytes, while smart contract headers are 124 bytes due to the inclusion of the storage hash and code hash. Although the State Trie does not store contract code or storage directly, it includes hashes that represent these components, ensuring the root of the State Trie encapsulates the entire state without actually containing all the data.

To execute a smart contract, nodes need its code. Although the theoretical code size limit is 24KB, the average size is around 1,630 bytes. Out of 7,953,851 deployed contracts, only 815,960 are unique, allowing for a 90\% reduction in storage needs by eliminating redundancies. When normalized for the number of accesses, the average contract size is 9,692 bytes, reflecting the greater complexity of widely used contracts.

Additionally, contracts maintain state through storage slots. Although the theoretical maximum storage is $2^{261}$ bytes per contract, the average is much lower at 1,888 bytes, and normalization reveals an effective size of 100.6MB due to transaction frequency. Transmitting full storage states across the network is impractical; an average transaction accesses only 9.45 slots, necessitating efficient data transfer strategies.

\subsection{Merkle Proofs}

Merkle proofs verify data validity, essential for confirming accounts and
storage slots. Proof sizes for accounts are stable, but vary widely for storage
slots, impacting the total data size transferred. Employing a logarithmic
regression model provided accurate proof size estimation, revealing that proofs
comprise 97\% of slot data. Since each transaction accesses 9.45 slots on average,
proof size is a significant contributor to total data size.

\subsection{Verkle Block Witness Size}
Verkle Tries are on the Ethereum roadmap to replace the current Merkle-Patricia Trie~\cite{EIP6800EthereumState}, mainly due to their significantly smaller proof size. This change will also alter the state's structure. Currently, Ethereum's structure consists of a main State Trie, with the storage root and code hash in the account header but not stored in the trie itself. Verkle Tries will create a unified trie where all data is stored together. Additionally, only the executed parts of a smart contract's code will need to be transmitted for validation, rather than the full code.

A key feature of Verkle Tries is their ability to efficiently create multi-proofs, enabling multiple values in the trie to be proven with a single proof, potentially reducing proof size per value by up to 90\%. Frequently accessed items are grouped in the same leaf, meaning the key into the trie is identical except for the last byte, resulting in minimal additional cost. The proof size difference between proving 1 value and 256 values within the same leaf is only 32 bytes.

We modeled the proof size based on the number of entries in the trie and the number of values proven, using existing data to estimate the proof size per block. While selective passing of code was not accounted for, we considered values grouped per leaf in our model. Due to multi-proofs' efficiency, the value of a single proof becomes irrelevant. Table \ref{tab:verkle} shows the projected witness size for a block.

\begin{table}
  \centering
\begin{tabular}{lll}
\toprule
 & Mean & Relative \\
\midrule
Account Header & 33.04 KB & 1.53\% \\
Storage & 71.69 KB & 3.32\% \\
Code & 1428.53 KB & 66.23\% \\
Verkle Witness & 623.57 KB & 28.91\% \\
\midrule
Total & 2156.83 KB & 100.00\% \\
\bottomrule
\end{tabular}
\caption{Distribution of Verkle block witness size, storage, code and account headers}
\label{tab:verkle}
\end{table}

\section{The State Network}\label{sec:statenetwork}

This section begins with an overview of our state network, followed by detailed explanations of key components, including account search, gossiping, caching, storage, and state synchronization. Additionally, we illustrate these concepts with an example of a transaction lifecycle.

\subsection{Overview}
In the state network, every node and data unit has an associated address or ID. Here, the data unit is the account, identified by its address. Although nodes also need addresses, these don't have to relate to any Ethereum address but must be of the same length.

A node retains an account if the first \( N \) bits of the account address match its address, where \( N \) is the prefix length. This structure aligns with the account tree, allowing each node to store a subtree from a specific path.

Nodes generate proofs efficiently by storing only hashes of branches they do not hold, needing just the \emph{proof path} to the subtree’s root. This avoids excessive storage, unlike storing random accounts, which would require full proofs for most accounts, leading to significant storage waste.

The network operates normally, with slight modifications to block proposals and gossiping processes. For block proposals, the proposer must ensure all necessary state data for transaction execution. While some state data is local, the proposer may need to request missing state data from the network, verify the proof, and proceed with the execution.

The gossiping process minimizes bandwidth use by not transmitting all state data. Instead, the sending node lists the required state but doesn't send the actual data. The receiving node identifies the missing state and requests it, along with its proof, from the sender. Nodes also maintain a cache to further optimize bandwidth.

\subsection{Account Search}
The core challenge is enabling a node to efficiently locate an account not stored locally. We employ a protocol inspired by Kademlia~\cite{maymounkovKademliaPeertoPeerInformation2002}, successfully used in systems like IPFS and BitTorrent for their resilience, adaptability, and customizability.

Each node maintains k-buckets to store information about other nodes, including addresses, IPs, ports, and prefix lengths. Entries are placed in buckets based on the length of the common prefix with the node's address. For instance, a node with address 010 and another with 011 would store the entry in bucket 2, as they share the first two bits. This ensures nodes have detailed knowledge of their closest peers, as detailed in Section \ref{kademlia}.

To search for an account, a node first checks its k-buckets for a node likely storing that account, identifiable via prefix lengths. If unsuccessful, the node searches for the address closest to the target. This step is necessary as Kademlia wasn’t designed for varying storage capacities; thus, proximity doesn’t always mean storage. A nearby node might have a long prefix length (small range), whereas a distant node might have a shorter prefix (broader range).

When a node receives an account request, it checks its storage and cache. To minimize bandwidth, requests can specify exact slots to return. A slot is a pairing of an address and value for contract state storage. If the node has the account, it returns the account, proof, requested slots or all slots (if unspecified), slot proofs, and contract code. 

If it lacks the account, it searches its k-buckets for nodes storing the account or closest to the address, returning the closest N nodes. The requester can then query these nodes. This redundancy ensures data availability; if one node fails or is slow, others can respond.

Due to the k-buckets' structure, nodes have detailed knowledge of their immediate vicinity, allowing the algorithm to converge logarithmically~\cite{maymounkovKademliaPeertoPeerInformation2002}.

\subsection{Gossiping}
After the initial block proposal, searching for an account is unnecessary. The block proposer gathers all required states, allowing efficient subsequent operations. When a node receives a block, it knows the sender has the state needed to validate it, eliminating the need for network-wide searches and distributing the load evenly.

Optimizing this process is crucial since only one node proposes a block, but all participants must receive and validate it. The block sender transmits the block along with a detailed list of the required state, specifying accessed accounts, storage slots, and code.

Sending the list instead of the state allows the receiver to determine what it already has cached. The sender, aware of the receiver's stored state based on address and prefix length, could omit known accounts, but this would require the receiver to reconstruct the list for further gossiping.

Sending the full list may slow the process, depending on network speed and cached state. However, the primary benefit is that the receiver can immediately propagate the block and list without waiting for the complete state and block execution, reducing network propagation delay.

Upon receiving a block, a node first propagates the block and state list. It then processes the list, searches its storage and cache, compiles a list of the missing state, and requests this from the block sender. Once received, the node verifies proofs and executes the block, updating its local storage and cache with the accessed state.

\subsection{Caching}
Effective caching is crucial for the network due to the high locality of state accesses and the bandwidth costs of the requesting state. 

There are three types of cache entries: account headers, slots, and code. 
Using these three types allows independent management of cache entries. Popular smart contracts often use many slots—on average, about 3,142,960 slots (equating to approximately 100.6 MB without proof size). However, each transaction accesses only about 9.45 slots. Indexes mapping addresses to specific values (e.g., account balance) mean that while accounts are frequently accessed, many slots are not. Managing accounts, slots, and code separately optimizes performance and resource usage without manual intervention.

A Most Frequently Used (MFU) caching strategy is optimal. It retains the most accessed items, as there is little time-related locality in state accesses but significant frequency-based locality. However, this strategy assumes nodes have processed preceding blocks or synced to the current state; otherwise, invalid states might be used during processing.

Keeping proofs up to date is critical, as modifications in accounts change the state root, invalidating existing proofs. Fortunately, if the state change proof is available, other proofs can be updated accordingly, though this process is resource-intensive. This issue also applies to slot proofs when another slot within the same account is modified.

To optimize, the proofs can be validated before executing a block. During execution, state changes and their proofs are tracked, and updates occur after executing the entire block, reducing computational overhead from constant proof updates.

\begin{figure*}[ht] % [htbp] specifies the placement of the figure (here, 'h' stands for "here," 't' for "top," 'b' for "bottom," and 'p' for "page")
    \centering % Center the figure horizontally
    % \vspace{-0pt} % Reduce space above the figure
    \includegraphics[width=\textwidth]{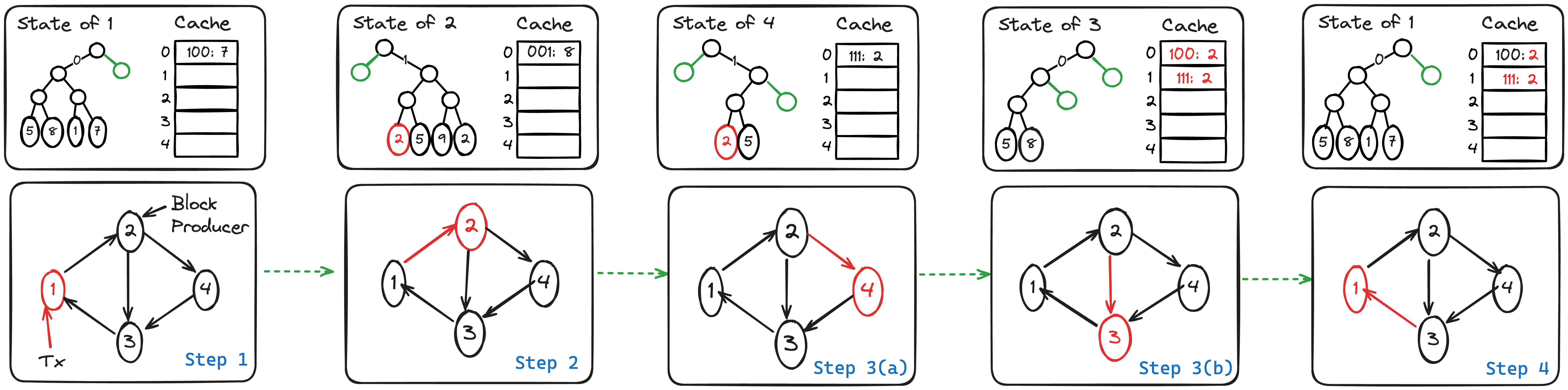}
    % \vspace{-15pt} % Reduce space below the figure
    \caption{Lifecycle of a transaction that changes address 100 to the value at address 111, which is 2.
    }
    \label{fig:txlifecycle}
\end{figure*}

\subsection{Storage}
Each node stores account data if the common prefix between the account address and the node address matches or exceeds the node's prefix length. This data is stored in a Merkle-Patricia tree, and the node maintains an up-to-date "proof path" leading to the tree's root. To construct a full proof for an account, the node combines this proof path with the proof it generates from its local tree.

Slots are stored in their own storage tree, while code is stored separately to facilitate reuse across multiple accounts.

\subsection{State Sync}
Efficiently syncing state when new nodes join, or old nodes rejoin is crucial to handling network churn effectively.

Thanks to the capability of operating in a stateless manner, new nodes can start functioning without the full state. They can store new changes while syncing the rest of the state iteratively to minimize cached data during synchronization.

The primary operation for state syncing is requesting a snapshot of the state at a specific path. This allows a node to obtain an up-to-date state view from that path, which it can use to build its local state tree. After loading the snapshot, the node applies any stored changes from recently received blocks. This iterative process helps synchronize with the network efficiently.

Rejoining nodes with an outdated state portion faces a more complex issue. These nodes can request the hash of a node and all its children at a specific Merkle tree point. Comparing these hashes with the local tree's hashes allows the identification of updated tree parts. By iteratively repeating this operation on branches with conflicting hashes, the node can pinpoint outdated accounts and request them individually.

\subsection{Example of Transaction Lifecycle}
To aid understanding, we illustrate a transaction lifecycle within a network of four nodes, as shown in Figure~\ref{fig:txlifecycle}. Each node's state is visualized via a tree structure. The figure displays cache entries, with new entries highlighted in red and existing ones in black.

In Ethereum, a transaction represents a state transition. We use integers to represent state tree values, and the transaction modifies the value at address 100 to the value at address 111, which is 2. Nodes need values at both addresses, 100 and 111, for execution. Even if 100 isn't directly changed, it must be included as modifying related values can alter state proofs.

In Step 1, a client submits a transaction to Node 1, which gossips the Tx to Node 2, the block producer. Node 2 creates a block, executes the transaction, updates storage in step 2, and gossips the block to Nodes 3 and 4, respectively, in step 3(a) and step 3(b).

Nodes 3 and 4 receive the block simultaneously. Node 4 has the value at address 100 in storage and 111 in cache (Step 3a), so it executes the block, updates storage and cache, and gossips the block to Node 3. 

Node 3, having received the block from Node 2, lacks relevant values in storage or cache (Step 3b). It requests state and proofs from Node 2, executes the block, updates its cache, and gossips the block to Node 1.

Node 1, with no relevant values in storage but address 100 cached (Step 4), requests state and proof for address 111 from Node 3. After updating its cache, Nodes process the block and gossip back to Node 2, stopping the process since Node 2 was the block proposer.

\section{Evaluation} \label{sec:eval}

% This text has often referenced the trade-offs necessary to reduce storage
% requirements for nodes. 
This section evaluates the protocol's efficiency and feasibility, providing quantitative results in network search efficiency, bandwidth, and latency to analyze trade-offs.

\subsection{Benchmark Data}
To produce representative results, we need to test our protocol on data that
accurately reflects real-world scenarios. For this, we can reuse the data from
Section~\ref{ethereum-access-patterns}. Provided a large enough time slice is used, it should yield illustrative results. The primary information needed for benchmarking includes which data is accessed and its size.

\subsection{Storage}\label{subsec:storage}
The storage space saved depends on the data availability solution and redundancy factor. Without any data availability solution and a $10^{-9}$ data loss probability in a 1000-node network, aggregate storage needs are reduced by 97.94\% compared to the baseline of full replication on every node.

\subsection{Network Search Efficiency}
The key metric for network search was the average number of iterations needed to find the required state, indicating how often the searching node had to query others before locating the value. The minimum iteration was 1 when the node merely queried a node already present in its k-buckets.

\begin{figure}[h]
    \centering
    \includegraphics[width=0.4\textwidth]{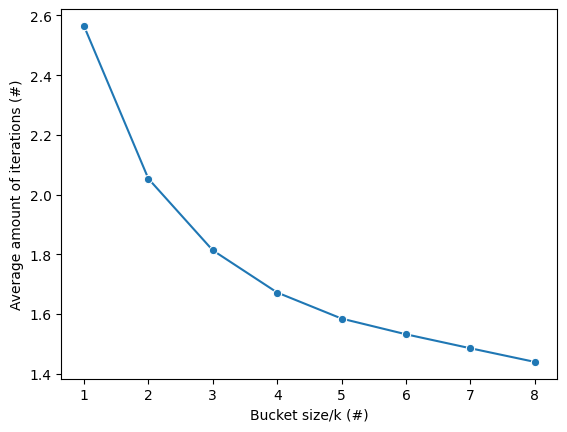}
    \caption{Effect of the bucket size (k) on the amount of iterations in state search}
    \label{fig:iterations-k}
\end{figure}

There was an almost perfect linear relationship between the average prefix length and the number of iterations needed to find the state. This linearity arose because Kademlia guaranteed a logarithmic number of iterations, while the number of items each node stored increased exponentially with the prefix length, balancing each other out.

Experiments showed that the network search efficiency was largely unaffected by the number of nodes, as the routing protocol relied on each node's k-buckets. Performance remained consistent if k-buckets remained unchanged. However, adjusting k-bucket sizes impacted the network search efficiency, as shown in Figure \ref{fig:iterations-k}. Increasing bucket size improved efficiency by reducing the number of iterations required due to the expanded knowledge each node gained about the network.

\subsection{Gossip Bandwidth}
We used \emph{Moka} for caching, a Rust caching library based on the TinyLFU algorithm~\cite{einzigerTinyLFUHighlyEfficient2017}.

Starting with empty caches would skew results, as nodes in operation never start empty. This issue worsens with larger cache sizes, taking longer to fill. To avoid this, we pre-initialized caches. We compiled and sorted by frequency a list of all states accessed during the observed period and inserted it into the cache until full, approximating the state of the MFU cache. While this introduces a bias towards frequently accessed states, Ethereum's high access locality makes this approach reasonable, as detailed in Section~\ref{ethereum-access-patterns}.

\begin{figure}[h]
    \centering
    \includegraphics[width=0.4\textwidth]{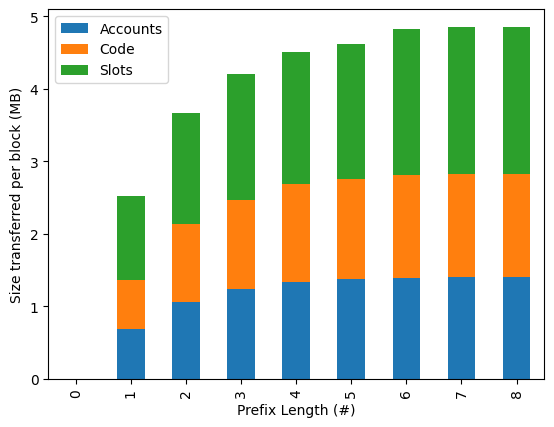}
    \caption{The average amount of additional bandwidth per block w.r.t. the
    prefix length of a node with no cache}
    \label{fig:bandwidth-prefix}
\end{figure}

The prefix length of a node significantly impacts additional bandwidth usage due to the storage-bandwidth trade-off. A node with a prefix length of zero stores everything, requiring no additional bandwidth. Figure~\ref{fig:bandwidth-prefix} shows that as the prefix length increases, additional bandwidth also increases since the node stores less state locally. Although the direct relationship between bandwidth and prefix length isn't immediately clear, comparing bandwidth to the proportion of state stored reveals a clear linear relationship. This aligns with the relationship \( R = 1 - 2^{\mathit{PL}} \), where \( R \) represents the proportion of state stored, and \( \mathit{PL} \) is the prefix length.

\begin{figure}[h]
    \centering
    \includegraphics[width=0.4\textwidth]{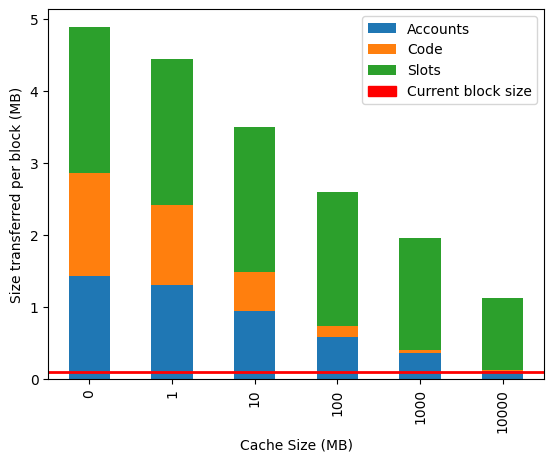}
    \caption{The average amount of additional bandwidth per block w.r.t. the size of the cache on a node with no fixed storage}
    \label{fig:bandwidth-cache}
\end{figure}

The second most important factor is the cache size, which directly impacts total bandwidth usage. We can see in Figure
\ref{fig:bandwidth-cache} that bandwidth drops quickly with small cache sizes but then shows diminishing returns, aligning with Section \ref{ethereum-access-patterns}. The red line in Figure \ref{fig:bandwidth-cache} shows the current size of a block. In full stateless
operation, without any cache, the amount of data transferred per block would
increase by a factor of almost 50 in comparison to the current data transferred
per block. 

It is interesting to observe how the proportions between the various data types change as the cache size increases. The code size diminishes quickly because bytecode tends to be referenced by multiple contracts, leading to relatively higher access frequency.

Even with a 10GB cache, a substantial amount of data still needs to be
transferred. This is intriguing, given that the total amount of state was
estimated to be around 50GB in 2020~\cite{SimplerEthereumSync2020}. The likely
cause is the significant overhead introduced by the proofs. This is particularly
problematic for slots, where only 3\% of the
data transferred is the actual slot data, while 97\% is the proof for the slot.

\begin{figure}[h]
    \centering
    \includegraphics[width=0.4\textwidth]{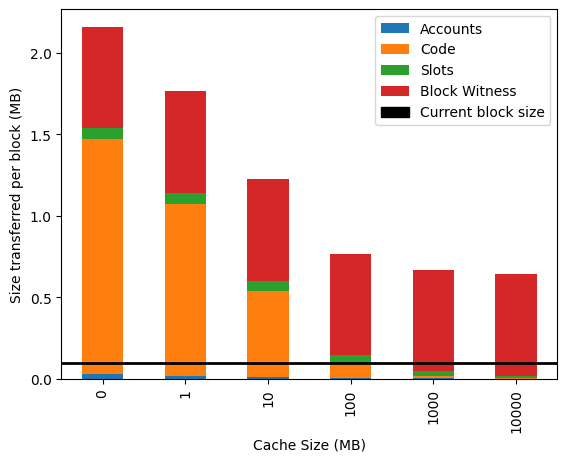}
    \caption{The average amount of additional bandwidth per block w.r.t. the size of the cache on a node with full Verkle block witness and no fixed storage}
    \label{fig:bandwidth-verkle}
\end{figure}

\subsubsection{Verkle Tree}

It is exactly due to this discrepancy that Ethereum is now moving away from
Merkle trees towards Verkle trees~\cite{EIP6800EthereumState}. While the
different tree structure makes it difficult to directly compare the two, it is
possible to make some estimation by considering the situation where each node
caches data, but the full block witness is sent together with each block. In
Figure \ref{fig:bandwidth-verkle}, we can see that the overall additional
bandwidth requirements decrease significantly. This change is attributed to two
factors: transferring less data due to the smaller proof size, and the ability
to cache more data due to the reduced storage size required for each cache entry
because the proof is not stored. With these adjustments, even a moderate cache
size of 100MB results in additional bandwidth usage that is in a similar order
of magnitude as the bandwidth currently used for block propagation.

\subsection{Latency}
Latency was a critical factor due to Ethereum's 12-second block target. If blocks were not propagated quickly enough, nodes might not have had sufficient time to process and vote on blocks. Failing to meet the 2/3 quorum within this window resulted in increased block times.

We used Simblock~\cite{aokiSimBlockBlockchainNetwork2019} to simulate block propagation and measure network latency. This modular and configurable simulator was ideal for researching block propagation in blockchain networks and was easily adaptable to our protocol.

Simblock did not allow modeling of the block itself to be sent before the rest of the state. Instead, we modeled an upper bound by combining the block and required state data into the block size. By adjusting the block size, we observed propagation behavior. Using the average Ethereum block size of 100kB (per Etherscan~\cite{etherscan.ioEthereumAverageBlock}) and a worst-case scenario of 5MB, we made comparative observations.

Another crucial parameter was the number of nodes in the network. Although this number fluctuated, a reasonable estimate was 7,000 nodes, as reported by Ethernodes~\cite{HistoryEthernodesOrg}.

\begin{figure}[h]
    \centering
    \includegraphics[width=0.4\textwidth]{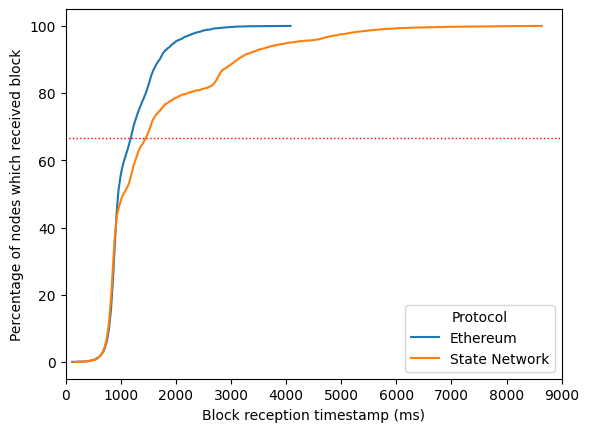}
    \caption{Comparison of block propagation between unmodified Ethereum and the state network}
    \label{fig:latency}
\end{figure}

In Figure \ref{fig:latency}, we observed that, despite a dramatic increase in block size, propagation latency changed little for most nodes. The time difference until the critical 2/3 of nodes received the block was approximately 270ms. Notably, the time for 95\% or more of the nodes to receive the block did increase significantly.

This latency increase under the worst-case 5MB block size scenario indicated potential challenges in maintaining the target 12-second block time, necessitating further optimizations in block propagation techniques.

\section{Related Work} \label{sec:relatedwork}

\subsection{Portal Network}

The Portal Network~\cite{PortalNetwork} uses a protocol similar to Kademlia for routing, but it has a different goal than our protocol. It does not focus on making it cheaper to run an Ethereum node but enables decentralized applications to get validated chain data without having to run an expensive Ethereum node or rely on an untrusted, centralized third party. It creates a totally different network to store the data, where \emph{bridge nodes} are both Ethereum full nodes and a Portal node that can push the new Ethereum data to the Portal network. The nodes then store part of the data such that the Network always has all data available, but each individual node only needs to store a small part. Compared to the Portal Network, our protocol is fully integrated into the underlying blockchain and enables transaction execution on low-resource nodes.

% The Portal Network aims to have each node expose the same JSON-RPC API as an Ethereum full node would. It relies on a variant of the discovery v5 protocol with the micro transport protocol instead of UDP to enable larger packets. 

% Portal Network has split the data of Ethereum in three categories, with each having their own sub-protocol: History for the block history, Beacon for the consensus layer and State for the state data.

% The Portal Network aims to have each node expose the same JSON-RPC API
% as an Ethereum full node would. It relies on a variant of the discovery
% v5 protocol with the micro transport protocol instead of UDP to enable
% larger packets. For routing, it uses a protocol similar to Kademlia.

% The Portal Network has a different goal. It does not focus on making it cheaper
% to run an Ethereum node, but enables decentralized applications to get validated
% chain data without having to run an expensive Ethereum node or rely on an
% untrusted, centralized third-party. Instead, it creates a totally different
% network to store the data, where \emph{bridge nodes} are both Ethereum full
% nodes and a Portal node which can push the new Ethereum data to the Portal
% network. The light clients then each store part of the data such that the
% Network always has all data available, but each individual node only needs to
% store a small part~\cite{PortalNetwork}.

% Portal Network has split the data of Ethereum in three categories, with each
% having their own sub-protocol: History for the block history, Beacon for the
% consensus layer and State for the state data.

\subsection{Low Resource Validators}\label{other-solutions}

% A significant amount of progress has been made surrounding light clients.
% Ethereum itself for example introduced the sync committee which made it faster
% for light clients to sync up with the consensus protocol~\cite{LightClients}. The
% use-cases for light clients until now have mostly been limited to validating
% data requested from other nodes for use in applications. 

There exist other solutions that
also attempt to enable low-resource validators in the blockchain but significantly change the way the
blockchain operates. In LightChain~\cite{hassanzadeh-nazarabadiLightChainScalableDHTBased2021}, everything is driven by the distributed hash table, this means that no traditional block gossiping is employed. They also use
a different consensus layer than the traditional Proof-of-Work or
Proof-of-Stake. Due to these reasons it can be a solution for new blockchains, but it is not possible to integrate into existing networks like ours.
LightChain replicates each piece of data on two nodes. In a network of 1000 nodes, 1 million transactions and 25,000 blocks, each node stores 30MB. The additional bandwidth necessary has not been estimated.

RemoteBlock~\cite{xiaRemoteBlockScalableBlockchain2024} is a system where nodes in a blockchain network have distinct roles; some are responsible for reaching consensus, while others manage the storage of blockchain data. This specialization reduces both the computational load on individual nodes and the overall storage demands of the network.

In contrast to RemoteBlock, our proposed approach specifies the routing protocol and allows every node in the network the opportunity to engage in all tasks - both storing data and participating in the consensus process. This maintains a more decentralized network structure, offering nodes the autonomy to decide their level of contribution based on their capacities. To improve scalability, we also streamline the gossip protocol to enhance communication efficiency and allow nodes to selectively determine the amount of data they wish to store, tailoring their involvement to their capabilities and resources. They do not provide a benchmark for the additional bandwidth required.

% Truthful Decentralized Blockchain Oracles\cite{blockChainOracles} is a stake-based, trustless protocol which allows nodes to vote on the valid response to a request. Thanks to the Merkle structure of many modern blockchain states, we can leverage a more lightweight approach by requiring the proof to be sent with the data. Note that we assume the node is up to date with the consensus layer, otherwise this would not be possible.

\subsection{Sharding}
Sharding was long on the Ethereum roadmap to increase the transaction throughput and alleviate storage requirements by splitting the network into distinct shards~\cite{KokorisKogias2018OmniLedgerAS, Zamani2018RapidChainSB}. Each node on the network is responsible only for the data within its specific shard. The biggest advantage of our protocol compared to sharding is that each node is able to fully choose its storage commitment while requiring less invasive changes to the blockchain protocol itself. However, our protocol does not increase the transaction throughput.

% (Proto-)Danksharding~\cite{Danksharding}, is the new alternative to sharding with the same goals, but makes use of L2 chains which can store blob data on the Ethereum chain temporarily for validation. As a result, they inherit the Ethereum security guarantees. While this approach will increase the total transaction throughput, it is improbable that this will cause the Ethereum blocks to not be full. As a result, the total state will still increase at the same rate.

\section{Limitations and Future Work}\label{sec:futurework}

\subsection{Data Availability Problem}
The current storage distribution system guarantees a replication factor probabilistically. However, this is not good enough in a blockchain system since the state getting lost would mean the blockchain is invalid. For this reason, there would need to be a system in place that guarantees a certain replication factor. There are already proposed solutions for this problem since it also exists in L2 chains~\cite{tasCryptoeconomicSecurityData2022}. The current system works probabilistically and the data loss probability is determined by the amount of nodes and the average ratio of data stored. As mentioned in \nameref{subsec:storage}, in a network with 1000 nodes and an average storage ratio of 2\%, the probability of data loss is $10^{-9}$.

\subsection{Verkle-optimized Adaptation}
The current protocol is optimized for a Merkle-Patricia Trie. While we have quantified the potential gains by moving to Verkle Tries, the protocol could be made more efficient by exploiting the properties of the new storage structure.

\section{Conclusion}
\label{cha:conclusion}
This paper addresses the pressing issue of blockchain validator nodes having to store an ever-increasing state. Increased storage costs drive validators and decentralized application developers towards centralized services, undermining network decentralization and security. We propose a state sharing protocol that distributes storage responsibility across the network, thus reducing storage costs for individual nodes. State lookups are enabled through a Kademlia-inspired routing protocol, enhancing block propagation, balancing network load, and retaining the overall transaction validation process without dramatically altering the blockchain's fundamental operation.

Our analysis shows the additional bandwidth required ranges from 1.5MB to 5MB per block, translating to 319GB -- 1,065GB of monthly bandwidth usage. While this increase is non-trivial, it remains within Ethereum’s 12-second block validation window, though it may incur additional costs for node operators. The introduction of Verkle Proofs can further reduce this bandwidth by more than 50\%. Our research indicates that decentralization and scalability in blockchain networks need not be at odds, and opens up avenues for future work focused on further optimizing bandwidth, reducing latency, and minimizing costs.

\section*{Acknowledgement}
This research is partially funded by the Research Fund KU Leuven, and by the Cybersecurity Research Program Flanders.

\bibliographystyle{IEEEtranN}
\bibliography{main.bib}

\vspace{12pt}

\end{document}